# Transverse profile expansion and homogenization at target for the injector Scheme-I test stand of China-ADS

YANG Zheng (杨征), TANG Jing-Yu (唐靖宇)*, YAN Fang (闫芳), GENG Hui-Ping (耿会平),PEI Shi-Lun (裴士伦), CHEN Yuan (陈沅)，LI Zhi-hui (李智慧),

Institute of High Energy Physics, Chinese Academy of Sciences, Beijing 100049, China

**Abstract:** For the injector Scheme-I test stand of the China-ADS, a beam with the maximum power of 100 kW will be produced and transported to the beam dump. At the beam dump, the beam power will be converted to thermal load and brought away by the cooling water. Two measures are taken to deal with the huge power density at the target. One is to enlarge the contact area between the beam and the target, and this is to be accomplished by expanding the beam profile at the target and using two copper plates each having a 20$^o$ inclination angle relative to the beam direction. The other is to produce more homogenous beam profile at the target to minimize the maximum power density. Here the beam dump line is designed to meet the requirement of beam expansion and homogenization, and the step-like field magnets are employed for the beam spot homogenization. The simulations results including space charge effects and errors show that the beam line can meet the requirements very well at the three different energies (3.2 MeV, 5 MeV and 10 MeV). In the meantime, the alternative beam design using standard multipole magnets is also presented.

**PACS:**

## 1. Introduction

The China-ADS (Accelerator Driven subcritical System) project is a strategic plan to solve the nuclear waste problem and the resource problems for the nuclear power plants [1].The driver accelerator runs in CW mode and accelerates the 10 mA proton beam to 1.5 GeV to bombard the target to produce high-flux neutrons, and Table 1 lists the main specifications. It can be seen that the ADS driver linac has very high beam power and very high reliability that are not possessed by any of the existing proton linacs in the world; innovative techniques must be applied. Among the R&D efforts, a test stand for the injector Scheme-I is being constructed at IHEP.

Table 1. Main specifications of the injector Scheme-I of China-ADS

| Parameters | Values |
|---|---|
| Final beam energy / MeV | 10 |
| Beam current / mA | 10 |
| Beam duty factor (%) | 100 |
| RF frequency / MHz | 325 |
| Beam power at target / kW | 100 |
| Beam energy at the RFQ exit / MeV | 3.2 |
| Beam energy at the 1$^{st}$ CM exit/ MeV | 5 |

The test stand of the injector Scheme-I has a similar configuration to the formal one for the China-ADS [1], with the main difference that the superconducting section is composed of two shorter cryomodules instead of a longer cryomodule in the formal design. It will be built and commissioned by steps. At the test stand, the accelerator will be able to produce a beam with the maximum power of 100 kW at 10 MeV. During the commissioning, the beam is transported to a high-power beam dump. By impinging the target at the beam dump, the beam power will be converted to thermal load and brought away by the cooling water. The target is composed by two copper plates weld together with 20$^o$ inclination angle, and then the contact area between the beam and target can be increased, hence the power density at the target surface can be lowered. Even so, it is still a big challenge if the 100 kW beam hits the target directly. Generally, the natural transverse beam distribution from a proton linac is more-or-less Gaussian, probably with a large beam halo. Similar to the other high beam power machines such as ESS [2] and CSNS [3], the dump beamline should be designed to expand and homogenize the transverse beam profile at the target.

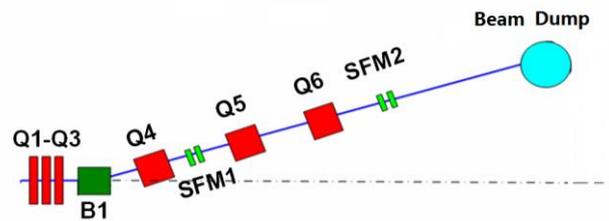

Fig. 1 Schematic layout of the beam dump line in the injector Scheme-I test stand of China-ADS.

## 2. Application of SFMs for beam spot homogenization

Non-linear magnets, such as the standard multipole magnets, the step-like field magnets (SFM) [4] and the special multipole magnets[5], are usually required for the beam spot homogenization at the target. Due to its excellent properties of low beam loss, compactness and low cost, the SFMs are adopted here. As given in Ref. [3], one SFM will produce an anti-symmetric field with almost zero field in the centre and two flat-top fields with sharp rise on the two sides. Two closely neighbouring SFMs with different designs will produce the required two-step field for the beam spot homogenization and for the beam halo control at the same time in one phase plane. This means that one needs at least two pairs of SFMs for the beam spot homogenization.

*Corresponding author: tangjy@ihep.ac.cn

The step-like field distribution of an SFM can be expressed approximately by

$$B(x) = \frac{F_s/L}{1+e^{-b(x-x_0)}} \quad (1)$$

where $L$ and $x_0$ are the effective length and the step position respectively, both are almost fixed in the magnet design. $F_s$ and $b$ are the field strength and the step sharpness, which can be changed by adjusting the power supplies of the two independent coils and are used for the optimization of the beam spot homogenization during commissioning.

## 3. Design of the dump beamline

According to the multiple-phase development plan of the test stand, in the first phase the commissioning is focused on the RFQ with the beam energy of 3.2 MeV; in the second phase it is focused on the first superconducting section of seven spoke cavities housed in the cryomodule CM1 with the beam energy of 5 MeV; in the third phase, it is focused on the second superconducting section which is identical to the first one and housed in the cryomodule CM2 with the beam energy of 10 MeV, as shown in Fig. 2. Therefore, the dump beamline has been designed to fit the beams with three different characteristics. As defined in the Ref. [1], the dump beamline is a part of MEBT2, but here it transports beams at the 3 different energies of 3.2 MeV (RFQ exit), 5 MeV (CM1 exit) and 10 MeV (CM2 exit), respectively.

### 3.1 Design goals and constraints

The dump beamline is designed with the following guidelines.

1) To facilitate the dump design which should be compact to fit in the existing tunnel, the transverse beam profile at the dump entrance should be more-or-less rectangular, the footprint (4-sigma×4-sigma) are required to be not smaller than 200×200 mm$^2$, 141×141 mm$^2$ and 110×110 mm$^2$ at 10 MeV, 5 MeV and 3.2 MeV respectively to make sure that the average beam power density in the transverse plane lower than 250 W/cm$^2$ at all the three energies. In the meantime, the peak power density is required to be less than 585 W/cm$^2$, which is determined by the 200 W/cm$^2$ power density limit at the copper target surface and the 20$^o$ inclination angle.

2) The homogeneity of the beam power density on the target should be as good as possible, which is determined by the design margin of the beam dump, about 10% in rms here.

3) At the beam dump entrance or so-called the target, the beam halo outside of ±130 mm in both the horizontal and vertical directions should be very low, e.g. less than 0.1% of the total beam power, as there is no cooling for the vacuum chamber which receives this part of the beam.

4) For the reasons of the hands-on maintenance and the device protection, the beam loss rate along the beamline should be controlled to be as low as possible. A total beam loss rate of a few tens Watts in the beamline is considered acceptable.

5) The beamline should be designed to adopt different beam characteristics at the three different stages. The whole beamline and the beam dump will be reinstalled during the later stages. The devices remain the same but the relative positions can be adjusted slightly.

Although the beam energy is low, a few quadrupoles with quite large apertures are applied to produce very flat beams at the positions of the SFMs. The optimizations are made to reduce the requirements on the quadruple strengths.

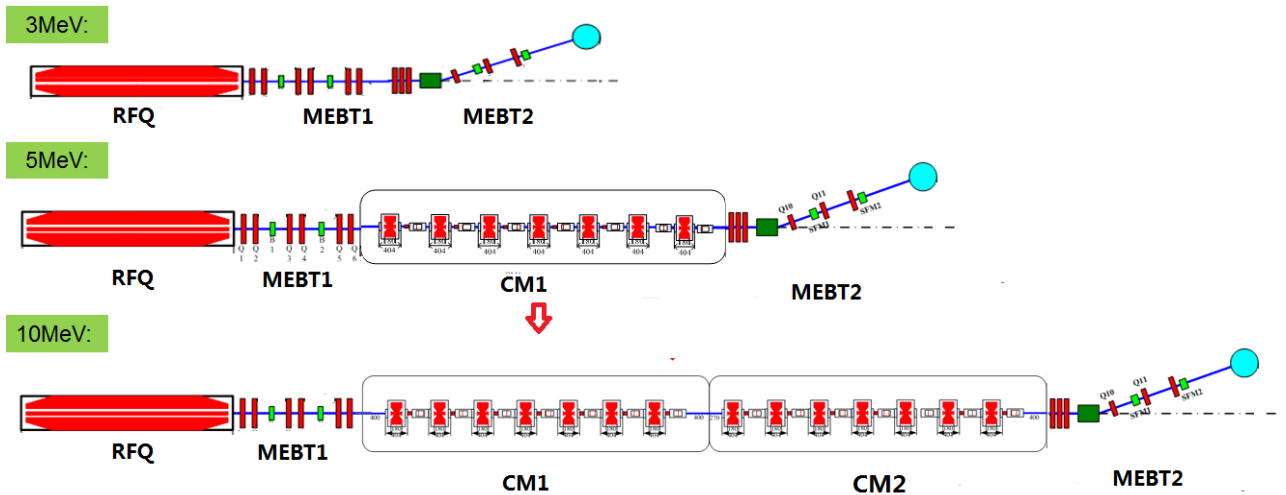

Fig. 2 Three stages of the construction and beam commissioning of the injector Scheme-I test stand of the China-ADS.

## 3.2 Design methods

As mentioned before, to control the beam homogeneity in both the horizontal and vertical planes, two pairs of SFMs are applied in the beamline. The aspect ratio of the beam cross-section at the SFMs should be large enough to decouple the nonlinear field in the two phase planes, e.g. larger than 6. At the same time, the phase advance between the SFMs and the target should be designed to be slightly different from $\pi$ or $2\pi$. Thus, to adjust the beam optics flexibly, one needs at least four quadruples, two of which are put before the 1$^{st}$ pair of SFMs, and the other two are located between the two pairs of SFMs.

A 15$^o$ bending magnet is introduced to avoid the back-streaming neutrons from the target into the cryostats. Therefore, two more quadrupoles are used for the transverse focusing, as shown in Fig. 1.

TRANSPORT [6], TURTLE[7] codes are used to set up the preliminary beam optics without applying the SFMs, while TRACEWIN [8] code is used to optimize the optics including space charge effects and the nonlinear fields of the SFMs. Usually, several iterations need to be carried out for the optimization of the linear optics and the SFMs.

For the multi-particle trackings by TRACEWIN, different initial beam distributions for 3.2 MeV, 5 MeV and 10 MeV at the beam line entrance are from the injector beam dynamics design, which are initially based on the RFQ design and contains 99072 macro-particles at the RFQ exit.

## 3.3 At 3.2MeV

At 3.2 MeV, the dump beamline is connected directly to MEBT1. Fig. 3 shows the beam phase space distribution at the MEBT1 exit. The linear beam optics for the dump beamline is shown in Fig. 4, which is designed to have flat beams at the SFMs and an enlarged beam profile at target with an initial emittance of 15 $\pi$mm.mrad which contains 99% particles.

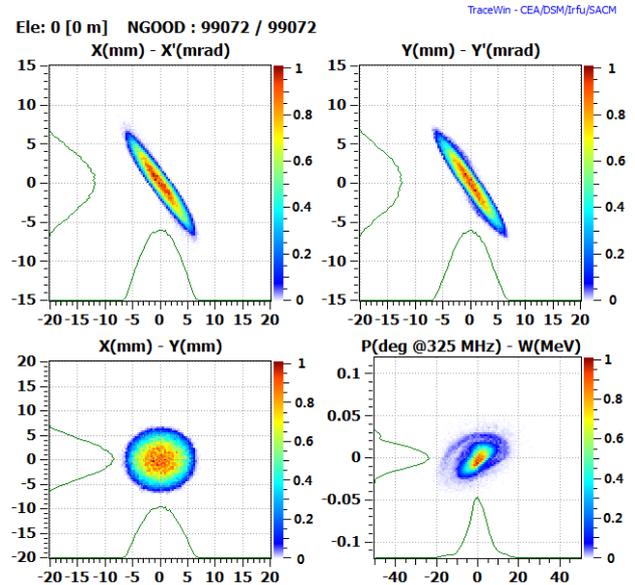

Fig. 3 Beam distributions in the phase spaces at the MEBT1 exit.

The multi-particle tracking result at the target entrance is shown in Fig. 5. For comparison, the design result by using octupoles in the places of the SFMs was also carried out and shown in the figure, whose homogeneity looks slightly worse at target. The mean square root homogeneity of the beam power density on the flat top is about 11% by using the SFMs, while it is about 19% for the case of octupoles. The maximum beam power density at target is 246 W/cm$^2$, which is far below the design requirement. With the given total macro-particles, there are statistical errors of a few percent in the homogeneity calculations that are due to relatively small numbers of macro-particles in each grid, and here 6×6 mm$^2$ size grids are used. The beam orbit correction, the error simulations show that no beam loss is observed for the case of SFMs.

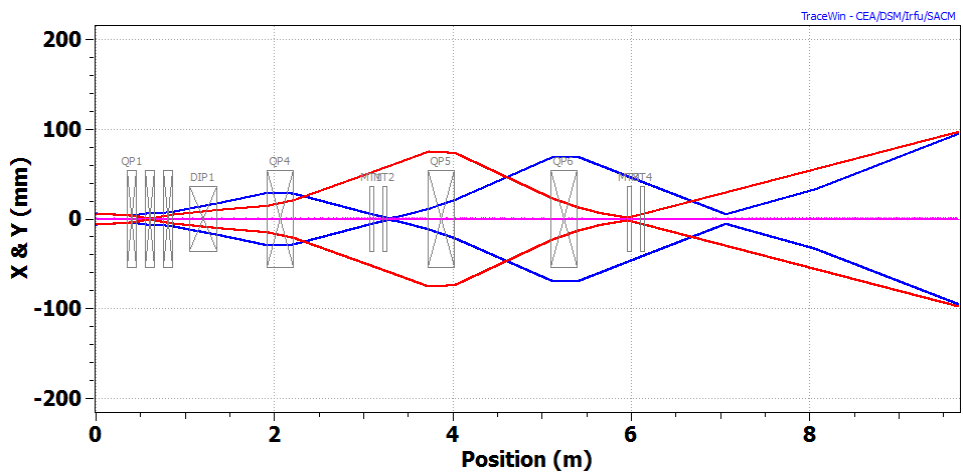

Fig. 4 Linear beam optics for the 3.2 MeV dump beamline.

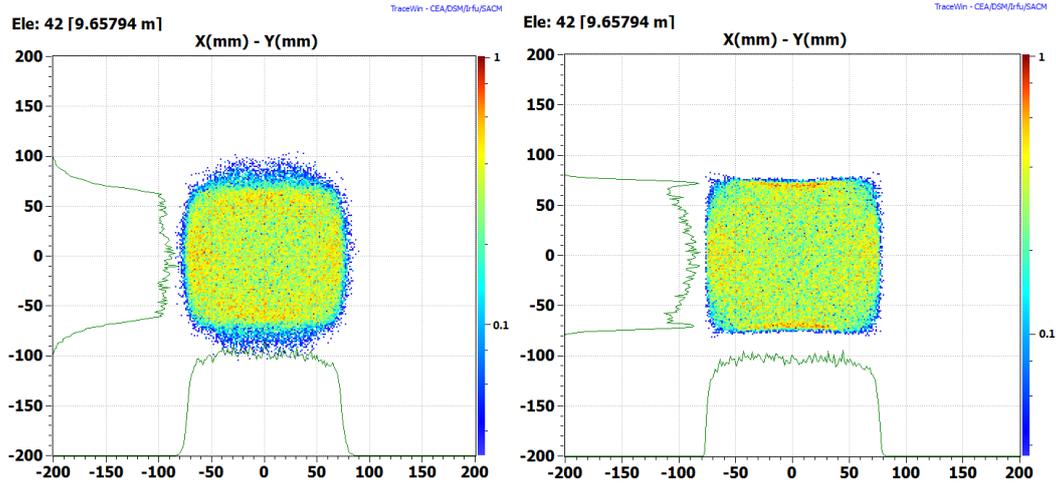

Fig.5 Transverse beam profiles at the beam dump entrance at 3.2 MeV (left: with SFMs; right: with octupoles)

### 3.4 At 5MeV and 10MeV

At higher beam energy, the beam emittance is smaller, while the required magnetic fields for the bending magnets, the quadrupole magnets and the SFMs are higher. In order to obtain a better control on the beam halo and the beam loss at both 5 MeV and 10 MeV, the $2^{nd}$ pair of SFMs is moved upstream by 0.3 m from the setup at 3.2 MeV.

Although they all meet the design goals, it is found that at 5 MeV it gives the best results on the beam halo control and the transverse profile homogeneity, compared with the cases at 3.2 MeV and 10 MeV. To control homogeneity error contributed by the statistical errors, the size of the grids are changed to 8×8 mm$^2$ at 5 MeV and 10×10 mm$^2$ at 10 MeV to keep averaged macro-particles per grid almost same for the three cases. The flat top homogeneity and the peak power of the beam power density at target are 11% in rms and 320 W/cm$^2$ at 5 MeV, 14% in rms and 385 W/cm$^2$ at 10 MeV. At 3.2 MeV, the performance is mainly limited by the relatively stronger space charge effect, while the main limitation for 10 MeV is the maximum magnetic field of the three large-aperture quadrupoles.

With the beam orbit correction, the average beam loss along the beam line is only several Watts at 5 MeV, and about 10 W at 10 MeV. This is considered acceptable at such low beam energy even though special treatments are required. Most of the beam loss happens between Q5 and Q6.

### 3.5 Simulations with 3D SFM field map

In the previous simulations, 2D SFM field maps are used. To check the performance with field maps as close to the reality as possible, 3D SFM field maps are generated by OPERA [9]. Fig. 6 shows the beam profile at the target at 10 MeV. It looks that the 3D result is only slightly worse than the 2D case, but still meet the design goal.

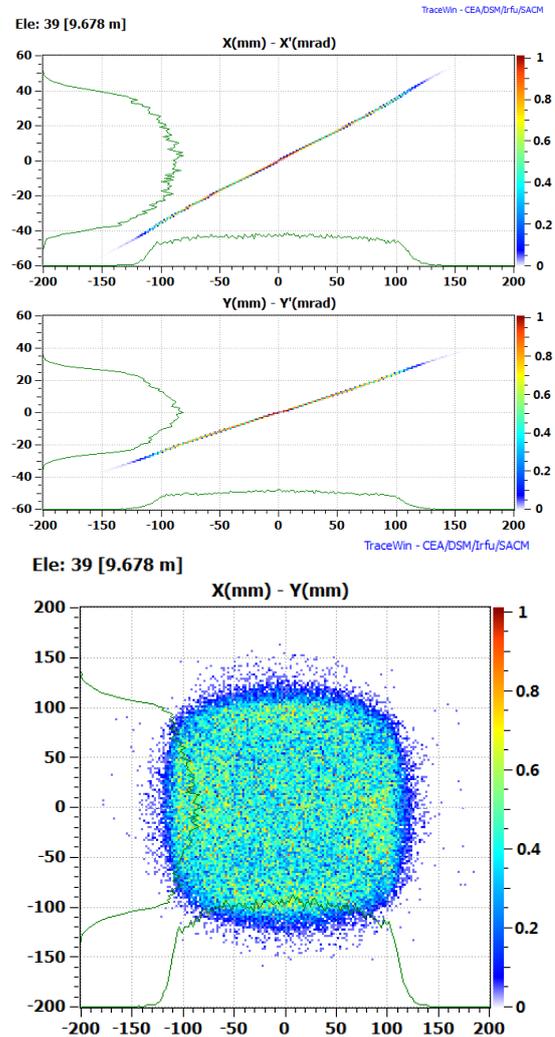

Fig 6 Particle distribution in the transverse phase planes at the target at 10 MeV with the 3D SFM field maps.

## 3.6 Summary of the designs

For all the 3 different energies, the beam losses along the beamline and the beam power distributions at the target are summarized in Table 2. The main design parameters for the magnets in the beamline are listed in Table 3.

Table 2 Beam losses along the beam line and the beam power distributions at the target

|  | 3-MeV | 5-MeV | 10-MeV |
|---|---|---|---|
| Loss without errors (W) | No loss | <2 | <2 |
| Loss with errors (W) | No loss | <5 | <10 |
| Major loss positions |  | DCCT, FCT | B1, Q5-Q6 |
| Beam power out of target (W) | 0 | 15 | 95 |
| Homogeneity across the flat top (%, rms) | 11 | 11 | 14 |
| Peak power density on target (W/cm$^2$) | 246 | 320 | 385 |

Table 3 Main design parameters for the magnets in the dump beamline

| Parameters | Length(m) | Magnet field gradient (T/m) /Strength (T) |
|---|---|---|
| Q1 | 0.1 | 0.98 |
| Q2 | 0.1 | 6.59 |
| Q3 | 0.1 | 11.30 |
| Q4 | 0.3 | 2.56 |
| SFMY-1 | 0.08 | 0.029 |
| SFMY-2 | 0.08 | 0.05 |
| Q5 | 0.3 | 1.76 |
| Q6 | 0.3 | 1.90 |
| SFMX-1 | 0.08 | 0.036 |
| SFMX-2 | 0.08 | 0.16 |

## 4. Conclusions

To facilitate the commissioning of the injector Scheme-I for the China-ADS linac, a dump beamline together with the beam dump is designed, which uses the SFMs to produce homogenized beam footprints at the target. The beam dynamics simulations show that the designed beamline can work very well at three different energies (3.2 MeV, 5 MeV and 10 MeV), which meets the requirements on the beam profile at target and the beam losses in the beamline.

*The authors would like to thank all the colleagues in the China-ADS beam dynamics group for very helpful discussions. The study was supported jointly by the CAS Strategic Priority Research Program- China-ADS and National Natural Science Foundation of China (Projects 11235012, 10975150).*